\documentclass[12pt]{iopart}
\usepackage[english]{babel}
\usepackage{hyperref}
\usepackage{graphicx}
\usepackage{caption}
\usepackage{pgf}	

\newcommand{\eqref}[1]{(\ref{#1})}

\begin{document}

\title{Subharmonic Shapiro steps of sliding colloidal
  monolayers in optical lattices}

\author{Stella V. Paronuzzi Ticco$^{1,2}$,   
  Gabriele Fornasier$^1$,
  Nicola Manini$^{1,2}$, 
  Giuseppe E. Santoro$^{2,3,4}$, Erio Tosatti$^{2,3,4}$,
  and Andrea Vanossi$^{2,3}$}
\address{$^1$
  Dipartimento di Fisica and CNR-INFM, Universit\`a di Milano,
  Via Celoria 16, 20133 Milano, Italy}
\address{$^2$
  International School for Advanced Studies (SISSA), Via Bonomea 265,
  I-34136 Trieste, Italy}
\address{$^3$
  CNR-IOM Democritos National Simulation Center,
  Via Bonomea 265, 34136 Trieste, Italy}
\address{$^4$ International Center for Theoretical Physics
  (ICTP), P.O.Box 586, I-34014 Trieste, Italy}
\ead{nicola.manini@fisica.unimi.it}

\begin{abstract}
%
%
  We investigate theoretically the possibility to observe dynamical mode
  locking, in the form of Shapiro steps, when a time-periodic potential
  or force modulation is applied to a two-dimensional (2D) lattice of
  colloidal particles that are dragged by an external force over an
  optically generated periodic potential.
  Here we present realistic molecular dynamics simulations of a 2D
  experimental setup, where the colloid sliding is realized through the
  motion of soliton lines between locally commensurate patches or
  domains, and where the Shapiro steps are predicted and analyzed.
  Interestingly, the jump between one step and the next is seen to
  correspond to a fixed number of colloids jumping from one patch to the
  next, across the soliton line boundary, during each AC cycle.
  In addition to ordinary ``integer'' steps, coinciding here with the
  synchronous rigid advancement of the whole colloid monolayer, our main
  prediction is the existence of additional smaller ``subharmonic'' steps
  due to localized solitonic regions of incommensurate layers executing
  synchronized slips, while the majority of the colloids remains pinned to
  a potential minimum.
  The current availability and wide parameter tunability of colloid
  monolayers makes these predictions potentially easy to access
  in an experimentally rich 2D geometrical configuration.
\end{abstract}


\maketitle

\section{Introduction}
Driven nonlinear systems can show widely intriguing patterns of varied and
often unexpected dynamical behavior.
An especially interesting case is that of many identical interacting and
crystallized particles moving in a periodic external potential: a common
situation, e.g.\ for adatom structures sliding on a crystalline surface.
In the special case when, besides the main time-independent force which
causes the sliding, an additional external "AC" force component is present
and oscillates periodically in time, synchronization phenomena are known to
occur, depending on the relative importance of the interactions and on the
geometrical arrangement of the particles.

Synchronization phenomena of this kind have long been described using
simple, low-dimensional, phenomenological models, which often suffice to
capture the main features of the complex dynamics involved.
The basic example is provided by the simple yet fundamental one-dimensional
(1D) Frenkel-Kontorova (FK) chain model
\cite{Frenkel38,Kontorova38a,Kontorova38b} whose soliton, or kink pattern
is known to exhibit a nontrivial intermittent dynamics, when additional
time-periodic forces act on the otherwise steadily sliding chain
\cite{Floria96}.
This simple 1D model however is not generally sufficient to describe in
full the complex dynamics of atoms at the interface of two materials in
relative motion.
Experimental studies of two-dimensional (2D) arrays of mutually repelling
colloids pushed across a periodic ``corrugation'' potential landscape have
recently been carried out, providing a closer analog of the sliding of
crystalline interfaces \cite{Mangold03,Bleil06,Bohlein12,Bohlein12PRL}.
Before sliding, the observed 2D patterns of misfit dislocation lines
(referred to as solitons or kinks) are very much akin to the Moir\'e
patterns formed by atomic overlayers over crystal surfaces.
Besides their surface physics analog, the colloid systems have an interest
on their own because they provide a ready parameter-controlled system for
the study of friction between crystalline surfaces \cite{Vanossi12,Bohlein12,
  Vanossi12PNAS}.

Here we explore, ahead of experiments, just very recently available in a
simpler 1D geometry \cite{Juniper15}, a situation where synchronization,
whence the existence and the actual dynamical nature of Shapiro steps,
is sought by oscillating in time either the pushing force or the
amplitude of the $(x,y)$ periodic corrugation potential.
In particular we focus on the second, corresponding to an ideal setup where
colloids thread a partially time-modulated spatially-periodic optical
potential, and investigate in particular the effects of the mismatch
between the colloid-colloid average spacing of the 2D colloid crystal and
the periodicity of the optical potential.

Shapiro steps were initially observed \cite{Shapiro63} in the context of
Josephson junctions.
The crucial ingredient at the start was Josephson's prediction
\cite{Josephson62} of a coherent current $I_c\sin{\phi}$ flowing between
two superconductors separated by a thin insulating barrier in presence of a
phase difference $\phi=\phi_1-\phi_2$ between the two superconductor order
parameters.
Moreover, a potential difference $V=\frac{\hbar}{2e}\frac{d\phi}{dt}$ was
predicted to be associated to a phase $\phi$ that changes in time, with the
remarkable consequence, known as AC Josephson effect, that the coherent
current would oscillate with a frequency $\omega_0=2eV_{\rm dc}/\hbar$ in
presence of a constant DC bias $V_{\rm dc}$.
The fingerprint of $\omega_0$ can be detected through an ingenious
resonance experiment \cite{Shapiro63}, in which a current $I(t)=I_{\rm
  dc}+I_{\rm ac}\sin(\omega t)$ is passed through the junction.
A simple model \cite{Stewart68,McCumber68} capturing resistive and
capacitive effects in the junction would allow us to write:
\begin{equation} \label{RCSJ_model:eqn}
  \frac{\hbar C}{2e} \frac{d^2 \phi}{dt^2} +
  \frac{\hbar}{2eR} \frac{d\phi}{dt} + I_{c} \sin\phi = I(t)
  \,,
\end{equation}
where $C$ is the capacitance of the junction, and $\frac{\hbar C}{2e}
\frac{d^2 \phi}{dt^2} = C\frac{dV}{dt}$ the corresponding current, while
$\frac{\hbar}{2eR} \frac{d\phi}{dt}=\frac{V}{R}$ accounts for the resistive
part of the current, due to quasi-particles.
In general, a $\phi(t)=\omega_0t + \Delta_{\phi}(t)$ where
$\Delta_{\phi}(t)$ is a bounded function, implies a (time-averaged) DC
voltage $V_{\rm dc}=\frac{\hbar\omega_0}{2e}$ across the junction.
In a non-resonant situation, $\sin{\phi(t)}$ oscillates in an aperiodic way
with zero average, $\langle{\sin{\phi(t)}}\rangle_{\rm time}=0$; hence the
(time-averaged) DC voltage predicted by Eq.~\eqref{RCSJ_model:eqn} follows
the usual Ohmic I-V characteristics, $V_{\rm dc}=RI_{\rm dc}$.
Resonances instead can make $\langle{\sin{\phi(t)}}\rangle_{\rm time} \neq
0$ by the following phase-locking mechanism \cite{Kautz96}: the bounded
variation $\Delta_{\phi}(t)$ becomes a periodic function of time,
characterized by the frequency $\omega$ of the external driving, while the
slope of $\phi(t)$ --- the frequency $\omega_0$ --- exactly matches a
multiple of the driving frequency $\omega$: $\omega_0=n\omega$.
In such resonant situations, $\sin{\phi(t)}$ becomes periodic and with
non-zero average, $\langle{\sin{\phi(t)}}\rangle_{\rm time} \neq 0$,
implying that the constant terms in Eq.~\eqref{RCSJ_model:eqn} obtained by
averaging over a period sum up to:
\begin{equation}
  \frac{V_{\rm dc}}{R} + I_c \langle{\sin{\phi(t)}}\rangle_{\rm time} =
  I_{\rm dc}
  \,.
\end{equation}
In essence, at resonances, $I_{\rm dc}$ can be changed by an amount $\sim
I_c \langle{\sin{\phi(t)}}\rangle_{\rm time}$ without affecting the voltage
bias $V_{\rm dc}$: these are the {\em Shapiro steps} in the I-V
characteristics, where the the voltage step is strictly quantized through a
finite range of current.

A mechanical analogy of the Josephson junction model of
Eq.~\eqref{RCSJ_model:eqn} is straightforward, with the phase difference
$\phi(t)$ playing the role of a coordinate $x(t)$, the voltage $V\propto
d\phi/dt$ that of the velocity of a particle moving in a washboard
potential, and the current $I(t)$ the role of an external force.
Shapiro steps have then a natural translation into ``quantized'' steps of
the velocity of the moving particle for a finite range of applied force,
when the washboard frequency goes in resonance with the AC perturbing
frequency $\omega$.
This resonance phenomenon is robust, and one could for instance
periodically modulate the amplitude of the space-periodic corrugation
potential [the analog of $I_c$ in Eq.~\eqref{RCSJ_model:eqn}] rather than
the external force, and the same physics would follow.

As is shown in Eq.~\eqref{RCSJ_model:eqn}, the Shapiro physics involves a
single degree of freedom.
However, in systems of many interacting particles, crystallized but not
rigid, synchronization phenomena of the Shapiro kind are likely to give
rise to additional nontrivial dynamical effects due to concerted
multi-particle motion.
The strong interest in synchronization phenomena in multiple physical
contexts has generated numerous reseach works in recent years, among
which we recall Refs.~\cite{Juniper15,Kvale91, Kolton01,Kokubo04,
  Eichberger10, Tekic09,Tekic10,Tekic11,Mu09,Mali12}.
The system where we propose to explore the Shapiro physics consists of a
collection of repulsive colloidal particles dragged by an external force
over a spatially periodic corrugation potential, whose amplitude includes
both a static and a time-oscillating part.
When the frequency $\omega$ of the oscillation is a multiple of the
characteristic (washboard) frequency $\omega_0$, the forced sliding motion
of the 2D colloid lattice can give rise to Shapiro steps.
Combined experimental and simulation studies already addressed in the past
the effect of synchronization and the ensuing Shapiro steps in Brownian
particle dynamics \cite{Mu09}.
Other simulation work \cite{Libal06} addressed the case where the 2D
sliding lattice and the periodic potential are fully matched, in
addition to ratcheting conditions in mixtures.
Very recently, the microscopic dynamics underlying mode locking in a
colloidal model system has been recorded in a simple, yet noteworthy, 1D
experimental setup \cite{Juniper15}.
In this work we study the feasibility of observing Shapiro steps under
reasonably realistic conditions in 2D sliding colloid monolayers, under the
wider range of conditions that can be realized experimentally.
More specifically, we will describe a planar system of monodisperse
repulsive particles, forming a 2D triangular lattice with spacing
$a_{\rm coll}$, sliding within a periodic corrugation potential due to
an optical lattice, also triangular with spacing $a_{\rm las}$, whose
amplitude is partly modulated periodically in time, under various
conditions as determined by different choices of the commensurability
ratio $\rho = a_{\rm las} / a_{\rm coll}$.
Using classical molecular dynamic we will monitor the individual motion
of the particles as well as that of their center of mass (CM) under the
action of a DC force.
As a function of this force the forward CM velocity of the colloidal
particles should, owing to Shapiro synchronization, become 'quantized' in
step-like structures as a function of the applied force.
Our aim is therefore to seek, detect and characterize these step
structures in relation to the absence (in the fully matched,
commensurate case $\rho=1$) or presence, e.g., for $\rho < 1$ of
pre-existing soliton-like defects or kinks due to mismatched relative
spacings of the colloidal lattice and of the periodic potential.
%

\section{The model}\label{model:sec}	

Molecular dynamics (MD) simulations are based on the same model already
introduced \cite{Vanossi12PNAS} to describe the motion of mutually
repulsive charged colloidal particles driven across a periodic potential
generated by a light interference pattern, as realized in experiments by
Bohlein {\it et al.} \cite{Bohlein12,Bohlein12PRL}.
In short, we describe the charged colloidal particles as classical
point-like objects, whose dynamics is affected by their mutual repulsion,
the action of external forces, plus the interaction with the viscous fluid
where they are immersed.

The equation of motion for the $j$-th particle is
\begin{equation}\label{motion}
  m\ddot{{\bf r}}_j (t) + \eta (\dot{{\bf r}}_j (t) - v_d \hat{{\bf x}}) =
  - \nabla_{{\bf r}_j}(U_2 + U_{\rm ext}) + {\bf f}_j (t) 
\,,
\end{equation} 
where ${\bf r}_j$ is the position of the $j$-th colloid in the 2D plane
where it can move, $\eta$ is the friction coefficient determined by the
effective viscosity of the fluid in which the colloids are immersed, and
$v_d$ is the fluid drift velocity, giving rise to a Stokes driving force
$F=\eta v_d$, pushing forward all colloidal particles.
$U_2$ is the two-body inter-particle potential; $U_{\rm ext}$ is the
corrugation potential describing the interaction with the spatially
periodic light-field pattern.
In this work we especially focus on the effect of time-periodic
oscillations of the amplitude of $U_{\rm ext}$.
The finite-temperature Brownian motion of colloids is introduced in a
standard Langevin approach, involving the viscous friction term $\eta
(\dot{{\bf r}}_j (t)~-~v_d \hat{{\bf x}})$, plus the Gaussian random
force ${\bf f}_j (t)$ \cite{Allen91}.
Due to the low overall velocity ($v_d \simeq 10~\mu$m\,s$^{-1}$) of the
colloidal particles, the inertial term can be neglected, and an
overdamped diffusive motion is reasonably assumed, with a sufficiently
large $\eta$.
In our work $\eta = 28$ expressed in the same model units described in
Ref.~\cite{Vanossi12PNAS}.
Typical simulation parameters are reported in Table~\ref{parameters:tab}.

\begin{table}
 \begin{center}
 \begin {tabular}{c|c|c|c|c|c|c|c}
 \hline
 $\rho$  & $\eta$ &$N$   & $N_{\rm minima}$&$Q$ &$\lambda_D$ & $L_x$ & $L_y$\\
 \hline
 $1$     &$28$    &$392$ & $392$     &$10^{13}$ &$0.03$  & $14$  & $14\sqrt3$\\
 $14/15$ &$28$    &$392$ & $450$     &$10^{13}$ &$0.03$  & $14$  & $14\sqrt3$\\
 $3/4$   &$28$    &$450$ & $800$     &$10^{13}$ &$0.03$  & $15$  & $15\sqrt3$\\
 \hline
 \end{tabular}
 \end{center}
 \caption{\label{parameters:tab}
   Adopted simulation parameters, expressed in model units
   \cite{Vanossi12PNAS}.
   For example, the simulation-cell sides $L_x$ and $L_y$ are given in
   units of the average colloid lattice spacing $a_{\rm coll}$.
   $N_{\rm minima}$ is the number of identical local minima of the
   corrugation function $W({\bf r})$ contained in a simulation cell.
 }
\end{table}

The two-body interaction potential is 
\begin{equation}\label{2-body_pot}
U_2 = \sum_{j<j'}^N V_{\rm Yuk}(|{\bf r}_j - {\bf r}_{j'}|)
\,,
\end{equation} 
where the screened Coulomb repulsion $V$ is a Yukawa potential:
\begin{equation}\label{coulomb-yukawa}
V_{\rm Yuk}(r) = \frac{Q}{r} {\rm exp}(-r/\lambda_D)
\,.
\end{equation}
The average nearest-neighbor separation of colloids in
Ref.~\cite{Bohlein12} is $r = a_{\rm coll} \simeq 5.7~\mu{\rm m} \simeq
30 \lambda_D$.
Under confinement, this repulsion establishes a triangular lattice of
colloids, which can be thought as reasonably defect free at temperatures
not too high and sizes sufficiently small to render the possibility of
Nelson-Halperin dislocations \cite{Nelson79} irrelevant to the present
case.

The one-body term
\begin{equation}\label{1-body_pot}
U_{\rm ext} = \sum_{j}^N V_{\rm ext}({\bf r}_j,t)
\,,
\end{equation}
describes the interaction of the colloids with the 2D spatially periodic
potential, representing the optical lattice corrugation.
Its actual form, determined by $V_{\rm ext}({\bf r})$, could in
principle be shaped with some freedom, but the simplest sinusoidal form
is sufficient to describe the main phenomena.
We thus take:
\begin{equation}\label{time_dep_pot}
V_{\rm ext}({\bf r},t) = \left[V_0 + \Delta_0 \sin(\omega t)\right] W
({\bf r})
\,,
\end{equation} 
where
\begin{eqnarray}\label{periodic_corrugation_expansion}
  W({\bf r})
  =
  -\frac 19 \, \left[ 3
    +  4
    \cos\left(\frac{2 \pi r_y}{\sqrt{3}\, a_{\rm las}}\right)
    \cos\left(\frac{2 \pi r_x}{a_{\rm las}}\right)
    + 2 \cos\left(\frac{4 \pi r_y}{\sqrt{3} \,a_{\rm las}}\right)
    \right]
,	
\end{eqnarray}
a unit-amplitude eggcarton-type periodic potential of $6$-fold symmetry
representing the optical-lattice triangular corrugation. 
The time-dependent amplitude $V_0 + \Delta_0 \sin(\omega t)$ gives to the
corrugation an additional sinusoidal time dependence that can induce
synchronization effects.
Its amplitude $2 \Delta_0$ yields a total corrugation ranging between
$V_0-\Delta_0$ and $V_0+\Delta_0$ with an AC modulation frequency $\nu =
\omega /( 2\pi)$.
The minimum force required to dislodge an isolated colloid at $T=0$
along direction $x$ is at time $t$
\begin{equation}\label{F_s1}
F_{s1} = \frac{8 \pi}9 \, \frac{V_0 + \Delta_0 \sin(\omega t)}{a_{\rm las}}
\,.
\end{equation}
The quantity $8 V_0/(9 a_{\rm las})$ is used as our unit of force.
In general, the spacing $a_{\rm las}$ of the corrugation lattice
potential $W(x,y)$ can be tuned in such a way to be either matched or
mismatched with the colloidal lattice.
In mismatched configurations, one observes, both in simulations
\cite{Vanossi12PNAS} and in experiments \cite{Bohlein12}, the formation
of misfit dislocations also called {\em topological solitons}, or
kinks/antikinks in the language of the FK model \cite{Braunbook}.
Their existence and their motion under the external force dominates the
frictional properties of the sliding lattice, since it provides the
mechanism for mass transport \cite{VanossiJPCM}.

As for the oscillation frequency, it needs to be low enough for the
overdamped colloidal system to be allowed enough time to follow
adiabatically the AC modulation in the corrugation-amplitude oscillation.
The specific condition is $\nu\ll \eta/m$, and we use $\nu<0.1$ in model
units.
Shapiro-step structures can arise out of synchronization of the washboard
frequency $\omega_0$ of the colloids driven by $F$ over the corrugation
potential, with the time-modulation frequency $\omega$ of
Eq.~\eqref{time_dep_pot}.
At finite temperature (e.g.\ in experimental conditions), $\nu$ must also
not to be so small that particle random diffusion overshadows the ordered
synchronized crossing of potential energy barrier.

Our protocol begins with a preliminary relaxation preparing the sample
in its equilibrium, force-free initial state.
Sliding simulations are then conducted by applying a force $F=\eta v_d$
along $x$ on each particle, similar to the viscous drag of the flowing
fluid in experiments.
The static force $F$ is slowly ramped upwards in small steps $\Delta F$,
so as to explore the desired force range.
At each force value, after an initial transient lasting between $1$ and
$10$ oscillation periods, the particle CM displacement $\Delta x_{\rm cm}$
is extracted for the successive simulation time $\tau$, typically a large
integer (of order $100$) number of periods $\nu^{-1}$ of the time-dependent
perturbation.
We then record the average speed $v_{\rm cm} = \Delta x_{\rm cm}/\tau$
for each value of the driving force $F$.

\begin{figure}
\centerline{
\includegraphics[width=0.4\textwidth,angle=0,clip=]{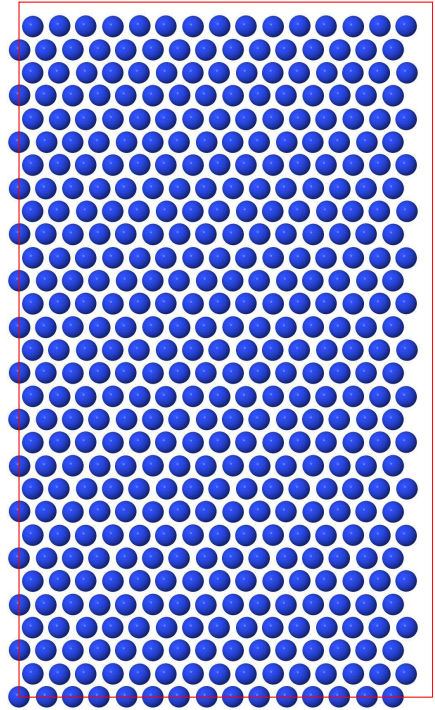}
}
\caption{\label{simulation_cell}
  The $L_x\times L_y = 15\,a_{\rm coll}\,\times\, 15 \sqrt3 \,a_{\rm
    coll}$ rectangular simulation cell (solid line) adopted for the
  $\rho=3/4$ simulations.
  In this initial snapshot, the $450$ particles still form a perfect
  triangular lattice.
}
\end{figure}

Experimental conditions \cite{Bohlein12} actually involve an inhomogeneous
2D configuration, whereby only a circular central portion of the colloid
monolayer raft is submitted to the laser field with its associated optical
lattice potential.
However, only phenomena in the central part of this circle are eventually
studied, and there the colloids are virtually homogeneous.
Therefore, we conduct the present study in a fully homogeneous 2D colloid
lattice with periodic boundary conditions (PBC) qualitatively represented
in Fig.~\ref{simulation_cell}.
Given a cell size and a particle number, we will simulate either the
commensurate and fully matched geometry ($\rho =1$), or mismatched
underdense geometries ($\rho < 1$), by tuning appropriately the corrugation
lattice spacing $a_{\rm las}$ relative to that of the colloid lattice,
$a_{\rm coll}$.

\section{Perfect lattice matching}\label{commensurate:sec}

\begin{figure}
\centerline{
\includegraphics[width=0.48\textwidth,angle=0,clip=]{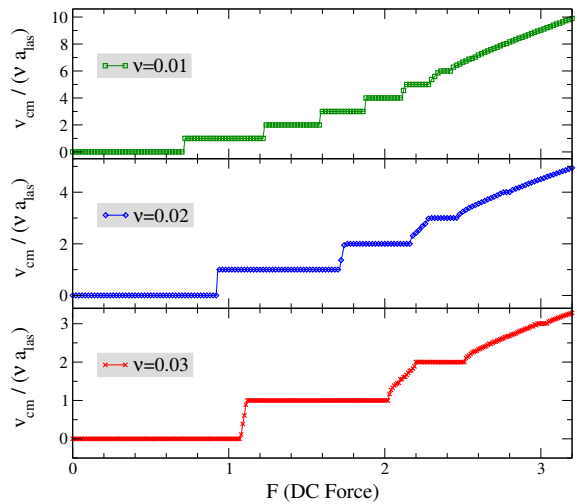}
}
\caption{\label{shapiro_commensurate}
$T=0$ Shapiro steps observed in the CM velocity parallel to the 
applied force obtained for a fully commensurate colloid lattice 
($a_{\rm las}=a_{\rm coll}$, $\rho=1$) with a corrugation amplitude 
$V_0=0.5$, a modulation amplitude $\Delta_0=0.2$, for values of the 
frequency $\nu=0.01$, $0.02$, and $0.03$ (in model units).
The velocity is scaled by $\nu a_{\rm las}$, to emphasize the
proportionality of the exactly quantized steps to the modulation
frequency.
}
\end{figure}

We begin with the fully matched commensurate geometry where, at $\rho =1$,
the two triangular lattices coincide, and no preliminary relaxation is
needed, each particle falling in a potential minimum.
We then turn on the sliding force with a time-dependent modulation of
amplitude $\Delta_0=40 \%\, V_0$, with an intermediate corrugation strength
$V_0 = 0.5$, at $T=0$.
The average CM velocity as a function of the external force $F$ is
presented in Fig.~\ref{shapiro_commensurate}.
As a function of the sliding force, it shows Shapiro steps, i.e.\ plateaus
in the average CM velocity.
These steps correspond to integer multiples of $\nu a_{\rm las}$ in the
force-velocity response.
In these $T=0$ simulations the motion of the colloids is deterministic,
leading to perfectly flat steps with null error bars, as long as the CM
speed is averaged over an integer number of periods $\nu^{-1}$.
The reference velocity $\nu a_{\rm las}$ is that of a particle advancing by
one corrugation spacing $a_{\rm las}$ in one oscillation period $\nu^{-1}$.
Depending on the applied DC force, all particles advance together by an
integer number of lattice spacings $a_{\rm las}$ at each period of the
AC modulation.
As a function of $\nu$ both the velocity jump between successive Shapiro
steps, and the force step width increase proportionally to $\nu$, as
expected.
Correspondingly, the number of steps in a given force range decreases
for increasing $\nu$.
In the large force regime, for $F>\frac{8 \pi}9 \, (V_0 + \Delta_0) /
a_{\rm las} \simeq 1.95$, the steps rapidly decrease in width, and
merge into a smooth velocity increase proportional to force,
characteristic of a viscous regime, similar to the non-modulated
$\Delta_0=0$ dynamics \cite{Bohlein12,Vanossi12PNAS}.

In this zero-temperature lattice-matched system, all colloids execute
the same movement to negotiate the crossing of the same energy barriers.
As a result, the colloid-colloid mutual spacing remains constant, and the
colloid-colloid internal force does no work.
The dynamics would be exactly the same if the colloidal particles were
not interacting, or if the simulation involved one particle only, moving
in the same potential-energy profile.
As a check, we repeated the simulations with a single particle, indeed 
recovering exactly the same patterns as in Fig.~\ref{shapiro_commensurate}.

\begin{figure}
\centerline{
\includegraphics[width=0.48\textwidth,angle=0,clip=]{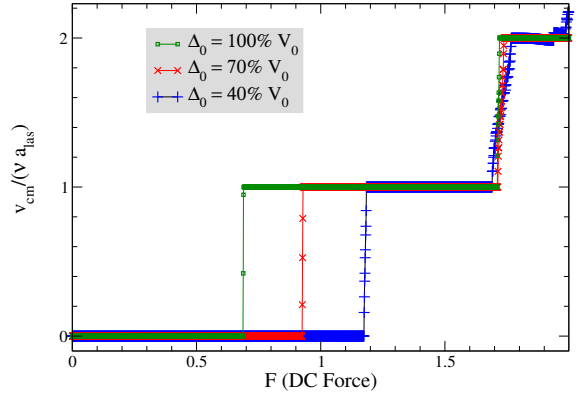}
}
\caption{\label{comm_V0.5}
  Shapiro steps for $\nu=0.02$, $V_0=0.5$, for increasing AC modulation
  amplitude $\Delta_0$.
  Pluses: $\Delta_0 = 40 \%\,V_0$, same as in the central panel of
  Fig.~\ref{shapiro_commensurate}.
  Crosses: $\Delta_0=70\% \,V_0$.
  Squares: $\Delta_0=100\% \,V_0$.
  The step heights are unchanged, their widths increase, and the jump
  between successive steps becomes steeper for increasing $\Delta_0$.
}
\end{figure}

We explore now how the  AC modulation amplitude $\Delta_0$ affects these
Shapiro steps.
Figure~\ref{comm_V0.5} shows the velocity curves obtained with the same
average corrugation amplitude ($V_0=0.5$) and AC modulation amplitude
rising from the same initial value $\Delta_0=40\% \,V_0$ as in the
central panel of Fig.~\ref{shapiro_commensurate} ($\nu=0.02$), to larger
values $\Delta_0=70\% \,V_0$ and $\Delta_0=100\% \,V_0$, at the same
frequency.
For larger $\Delta_0$ the colloid system starts to move earlier, i.e., at a
smaller DC force than for a smaller $\Delta_0$.
Note that at $100 \%$ relative modulation amplitude, every oscillation
period has a brief instant where the colloid layer actually slides freely.
Despite this free-sliding instant, the force needed for the monolayer to
depin and reach the first Shapiro step, even though smaller than at $40\%$
modulation amplitude, appears anyway to be nonzero.
The reason for that is the finite time it takes the colloid to drift across
the momentarily turned off barrier to the next substrate minimum, when the
driving force $F$ is small.
This issue is discussed in detail in Sect.~\ref{depinning:sec} below.

\subsection{Thermal effects}

\begin{figure}
\centerline{
\includegraphics[width=0.48\textwidth,angle=0,clip=]{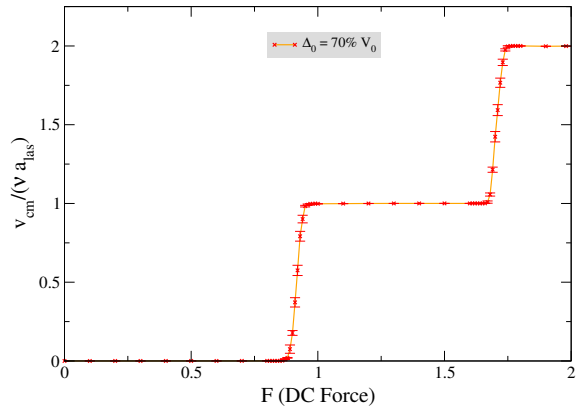}
}
\caption{\label{T_comm_0.04}
  Finite-temperature Shapiro-step structure for a fully commensurate
  system.
  Temperature $T=T_{\rm room}$, and the other simulation conditions
  ($\nu=0.02$, $V_0=0.5$, $\Delta_0=70\% \,V_0$) are the same as for the
  crosses data of Fig.~\ref{comm_V0.5}.
  The effect of thermal motion is barely detectable, apart from a slight
  rounding on the step edges.
  Thermal fluctuations of the CM velocity, reflected by the error bars,
  are visible at rises between steps, but negligible inside the steps.
}
\end{figure}

So far we worked at $T=0$, because that regime yields a clearer picture.
Next, however, we can ask how the Shapiro steps might be affected by
thermal fluctuations in the experimental, room-temperature conditions.
To address this question, we perform new simulations at room temperature
$T=T_{\rm room}$ ($T_{\rm room} = 0.04$ in system units), still at $\rho=1$
and keeping all other parameters (mimicking the experimental colloid system
of Ref.~\cite{Bohlein12}) the same.
Figure~\ref{T_comm_0.04} shows that the Shapiro-step structure is barely
affected; $T_{\rm room}$ is a very moderate temperature for this system.
Thermal fluctuations (estimated by the standard deviation of $v_{\rm cm}$
obtained over several sliding simulations realized with independent
random-number sequences and different initial configurations ensuring a
better sampling of the thermal equilibrium state) are visible, as reflected
by the error bars, at jumps between steps.
Thermal fluctuations produce a rounding of the plateau onset, but are
barely visible inside each plateau.

\section{Mismatched lattice spacing}\label{incommensurate:sec}

When the commensurability ratio
$\rho=a_{\rm las}/a_{\rm coll}$
differs from unity the colloid and the corrugation lattices are mismatched.
As we shall see, in this case the internal structure of the sliding lattice
plays a role, and the expected Shapiro-step phenomena become more
interesting.
It must be clarified at the outset that the Shapiro steps exist only if the
system is pinned by a finite static friction force.
For example, a 2D colloid lattice incommensurate with a weak periodic
corrugation is an unpinned system which can slide ``superlubrically''
without static friction, and will exhibit no Shapiro steps.
The same incommensurate lattice must however become pinned when the
corrugation amplitude exceeds a first-order depinning-pinning transition
threshold value \cite{MandelliPRB15}, and here Shapiro steps can exist.
This is the situation which we concentrate upon here.
We describe in detail two different examples of mismatched configurations,
$\rho=14/15\simeq 0.93$ and $\rho=3/4= 0.75$.
Both are \emph{underdense} systems compared to the perfectly matched one
with $\rho = 1$ considered in Sec.~\ref{commensurate:sec}, and both are of
course rationally commensurate.
However, the commensurabilty of $\rho=3/4$ is so to speak stronger, that in
$\rho=14/15$ weaker, the latter case in a sense closer to incommensurability,
itself a condition impossible to reach in a finite-size PBC realization.
We will further assume that the colloids and the corrugation lattice do
not undergo a relative rotation, e.g., of the type studied in
Refs.~\cite{Novaco77,McTague79,Shiba79,Shiba80,MandelliPRL15}, and
therefore remain aligned.
With that stipulation we adopt for $\rho=14/15$ a rectangular periodic
simulation cell involving $392$ colloids threading $450$ corrugation
potential minima.
With $14$ colloidal particles threading 15 potential wells along each
principal direction, we obtain a widely spaced hexagonal network of misfit
dislocation lines ({\em antisolitons}), where vacancies segregate.
Details of this case will be shown later.
For $\rho=3/4$ we also adopt a rectangular supercell, here involving $450$
colloids and $800$ potential minima, shown in Fig.~\ref{simulation_cell}.
The higher vacancy concentration leads to a greater mismatch and therefore
a denser packing of antisolitons.

\begin{figure}
\centerline{
\includegraphics[width=0.48\textwidth,angle=0,clip=]{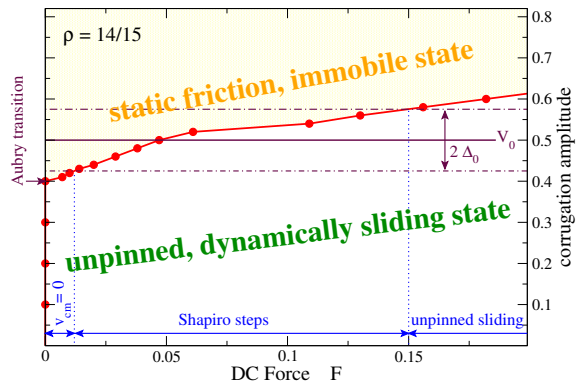}
}
\caption{\label{aubry_explanation}
DC Force $F$ -- corrugation amplitude $V_0$ phase diagram for
$\rho=14/15$ mismatched colloids, in the absence of AC modulation
($\Delta_0=0$).
The red dots and line represent the depinning
line where at $T=0$ the colloid abandons the statically pinned state (shaded yellow region)
in favor of a (nearly) free-sliding state for larger force $F$.
A truly incommensurate monolayer, where $\rho$ is irrational, would
exhibit a very similar transition line, the 2D analogue of the 1D Aubry
transition \cite{MandelliPRB15, AubryLeDaeron}.
Horizontal solid and dot-dashed lines: a choice of corrugation ($V_0 =
0.5$) and AC modulation $\Delta_0 = 15\%\,V_0$ amplitudes.
In the force range for which both dot-dashed lines are in the shaded
region, the colloid layer is pinned.
For large force, when both dot-dashed lines lie below the depinning
curve, the colloids advance freely.
The intermediate force interval is the one affected by the
synchronization to the AC potential modulation, giving rise to Shapiro
steps.
}
\end{figure}

In these mismatched systems we must first of all identify the corrugation
amplitude at which, as was mentioned above, a pinning/depinning transition
takes place in the absence of AC modulation ($\Delta_0=0$).
In the limit of infinitesimal applied force, this transition was
recently discovered and characterized by Mandelli {\it et
  al.}~\cite{MandelliPRB15}.
Ignoring here the weak $14/15$ commensurability and taking this to
represent a truly incommensurate case, this is the 2D analog of the
celebrated 1D ``Aubry transition''
\cite{Braunbook,Peyrard83,MandelliPRB15} here taking place at $V_{0\,
  \rm crit} \simeq 0.4$ \cite{AubryRational:note}.
With this stipulation, for every corrugation value $V_0<V_{0\, \rm
  crit}$ the colloid sliding over the incommensurate corrugation is
``superlubric'', and will take place for arbitrarily small applied
force.
By contrast, when $V_0 > V_{0\, \rm crit} $, the monolayer is pinned,
and a finite static friction force $F > 0$ is required to make the
system slide \cite{MandelliPRB15}.
The curve of Fig.~\ref{aubry_explanation} represents precisely this
depinning transition line for the $\rho=14/15$ mismatched system, for
nonzero and increasing force.

To understand what happens when we superpose an AC corrugation
modulation $\Delta_0>0$ to the ``phase diagram'' of
Fig.~\ref{aubry_explanation}, we add horizontal dot-dashed lines
representing the maximum and minimum value covered by the overall
corrugation magnitude during an oscillation.
As an example, in Fig.~\ref{aubry_explanation} we sketch $V_0 = 0.5$
(horizontal solid line) and $ \Delta_0 = 15\%\,V_0$ 
(horizontal dot-dashed lines).
Synchronization between the washboard sliding frequency and the AC
modulation, and consequently Shapiro steps, can arise naturally in the
intermediate range of forces $0.02\leq F\leq 0.15$, where at each
modulation period the system crosses twice the pinned-unpinned
transition curve.
Figure~\ref{aubry_explanation} shows that depending on $V_0$ and
$\Delta_0$, the $F$ dependence can vary quite substantially.
For example, if $V_0+\Delta_0<V_{0\,\rm crit}$, then for any $F$ the
colloids will slide freely.
For $V_0+\Delta_0>V_{0\,\rm crit}$ but $V_0-\Delta_0<V_{0\,\rm crit}$
the initial pinned region ($F\leq 0.02$ in figure) would disappear: in
this case the first Shapiro step could extend to arbitrarily small
applied force, by reducing the modulation frequency $\nu$, as further
discussed in Sec.~\ref{depinning:sec}.
This phase diagram can be taken as a guide for the choice of the simulation
parameters.
In order to observe Shapiro steps, the colloid lattice must be pinned,
e.g., $V_0$ must be above the critical corrugation $V_{0\,\rm crit}$.
The range of force where Shapiro steps exist widens out as the
modulation amplitude $\Delta_0$ is increased.
The phase diagram for $\rho=3/4$ is quite similar to that of
$\rho=14/15$, and in particular $V_{0,\rm crit}\simeq 0.42$ is nearly
the same.

We now explore and describe the existence and nature of Shapiro steps in
these mismatched cases, adopting an arbitrary but reasonable corrugation
$V_0=1 > V_{0,\rm crit}$ and AC modulation $\Delta_0=60\%\,V_0$
(different parameters were explored too).
Unlike the $\rho=1$ case, here a preliminary and careful relaxation of
the initial force-free, equilibrium structure is required.
An initial simulation for a duration $\tau = 1000$ is thus performed
with a time-independent corrugation ($V_0$ turned on, $\Delta_0=0$), and
no external force applied ($F=0$).
Starting from the relaxed configuration, a sequence of force-driven
sliding simulations is run, now with the AC corrugation modulation
turned on.

\subsection{$\rho=14/15$}\label{14_15}

\subsubsection{Step structure.}

\begin{figure}
\centerline{
\includegraphics[width=0.48\textwidth,angle=0,clip=]{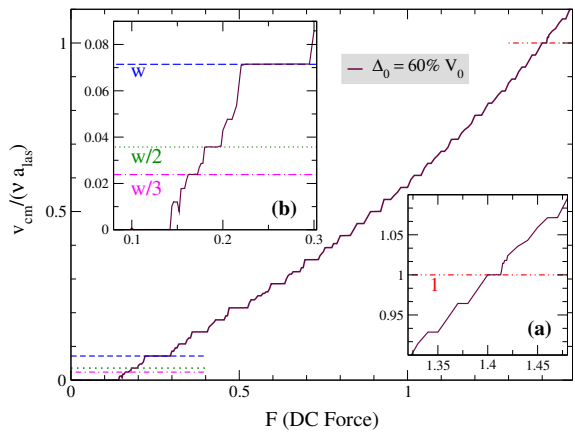}
}
\caption{\label{steps0.93}
  The Shapiro-step structure of the lattice mismatched $\rho= 14/15$
  configuration simulated at $T=0$, with corrugation amplitude $V_0=1$,
  modulation $\Delta_0 = 60\% \,V_0$, and AC frequency $\nu=0.02$.
  (a) A zoom-in of the region of the ``trivial'' commensurate step
  $v_{\rm cm}/(\nu\, a_{\rm las})=1$ corresponding to all colloids
  advancing by $a_{\rm las}$ at each period.
  (b) A zoom-in of the region of the subharmonic steps, highlighting
  those matching $w = |1-\rho^{-1}|=1/14$ and integer submultiples
  thereof.
}
\end{figure}

Let us consider $\rho < 1$, where it is useful to introduce as a misfit
measure the quantity
\begin{equation}\label{w_eq}
  w = \left|1-\frac{a_{\rm coll}}{a_{\rm las}}\right|
  = \left|1-\frac 1f \right|
  \,,
\end{equation} 
This quantity was found earlier to be the relevant parameter in the
context of simulations of mismatched crystalline systems in mutual
contact,
Refs.~\cite{Vanossi06,Manini07PRE,Manini07extended,Santoro06,Cesaratto07,Vanossi07Hyst,Vanossi07PRL,Vanossi08TribInt,Manini08Erice,Castelli08Lyon,Castelli09,Vigentini14}.
The results of sliding simulation for $w=1/14$, corresponding to $\rho =
14/15$, are shown in Fig.~\ref{steps0.93}.
The Shapiro steps exist here too,
%
but the mismatched geometry exhibits a far richer pattern of
Shapiro steps than the lattice-matched case: we detect tens of steps in
a much narrower force range than in the commensurate case of
Fig.~\ref{shapiro_commensurate}.

First off, the regular integer plateau at $v_{\rm cm} / (\nu\, a_{\rm
  las})=1$, at the same speed as the first step in fully matched
geometry, Fig.~\ref{comm_V0.5}, is detailed in the inset
Fig.~\ref{steps0.93}a.
Additionally, Fig.~\ref{steps0.93} exhibits several other subharmonic
steps at smaller driving force.
It has long been known that for mismatched lattices also {\em fractional
  subharmonic Shapiro plateaus}, i.e.\ plateau velocities which are not
integer multiples of $\nu\, a_{\rm las}$ \cite{Floria96,Falo93} are to
be expected.
In particular the inset Fig.~\ref{steps0.93}b zooms around a much lower
speed, given by 
\begin{equation}\label{quant}
\frac{v_{\rm cm}}{\nu\,a_{\rm las}} = w \;.
\end{equation} 
Here $w=1/14$ and velocity steps appear at $wm/n$ for several $m$ and
$n$ values.
All these $v_{\rm cm}<\nu\,a_{\rm las}$ plateaus are subharmonic Shapiro
steps.
Unlike in the integer plateaus, where all $N$ particles advance by one (or
several) lattice spacing(s) $a_{\rm las}$ in one period $\nu^{-1}$, at a
fractional plateau characterized by $v_{\rm cm}/(\nu\,a_{\rm las})= w\,m/n$,
a total fraction $mwN$ of all colloids advances by one lattice spacings
during $n$ modulation periods.

The quantization of sliding velocity to the value of Eq.~\eqref{quant}
was discovered and studied in the context of the sliding of mismatched
crystalline systems in mutual contact, and reported in the study of two
sliding surfaces with an atomically thin solid lubricant layer in
between \cite{Vanossi06,Manini07PRE,
  Manini07extended,Santoro06,Cesaratto07,Vanossi07Hyst,
  Vanossi07PRL,Vanossi08TribInt, Manini08Erice,Castelli08Lyon,
  Castelli09,Vigentini14}.
There, the ratio $\rho$ was that of the lattice spacing of a ``bottom''
slider to that of the lubricant layer, $w$ was the mean velocity at
which the lubricant moved over the bottom slider, in units of the
externally imposed speed of the ``top'' slider.
The sliding of the lubricant with relative speed $w$ arose in that case
because the vacancies (or interstitials) forming the antisolitons
(solitons) were the real entities dragged forward at the full speed of
the top layer, all (other) particles remaining essentially static.
In the following we shall clarify the relationship between the driven
soliton dynamics in the sliding-friction models and the subharmonic
steps in the dynamics of colloids driven in a modulated-amplitude
potential.

\subsubsection{Subharmonically sliding antisolitons.}
\label{Subharmonic Antisolitons}

\begin{figure}[t]
\centerline{
\includegraphics[width=0.48\textwidth,angle=0,clip=]{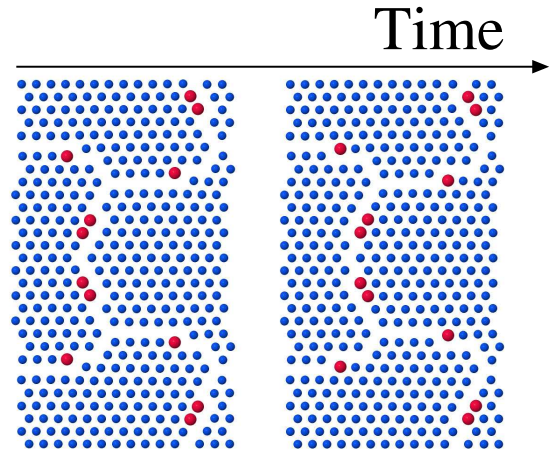}
}
\caption{\label{antisoliton_passage}
  $\rho=14/15$, $T=0$,  $w/3$ fractional plateau.
  The antisoliton pattern divides the colloid lattice into domains or
  patches inside which colloids are approximately matched to the static
  corrugation lattice.
  The underdense colloid monolayer slides rightward due to the leftward
  motion of the antisoliton line, while particles inside the domains do
  not slide.
  The specificity of this plateau is the behavior of particles at the
  antisoliton boundary.
  Left and right are two successive snapshots separated by a single
  oscillation period $\nu^{-1}$.
  At the end of this time interval, most colloids still occupy their
  initial positions, but the 12 highlighted particles (red) crossed the
  antisoliton line to join the next commensurate domain to the right.
  In the two following $\nu^{-1}$ periods the remaining $Nw - 12 = 16$
  particles will also cross, so that after the time $ n\nu^{-1}$ (here
  with $n=3$) all $Nw = 28$ particles will have crossed.
  Different plateaus with different $w$ and different subharmonic $n$
  will have an analogous but different pattern of traversing particles.
}
\end{figure}

\begin{figure}
\centerline{
\includegraphics[width=0.48\textwidth,angle=0,clip=]{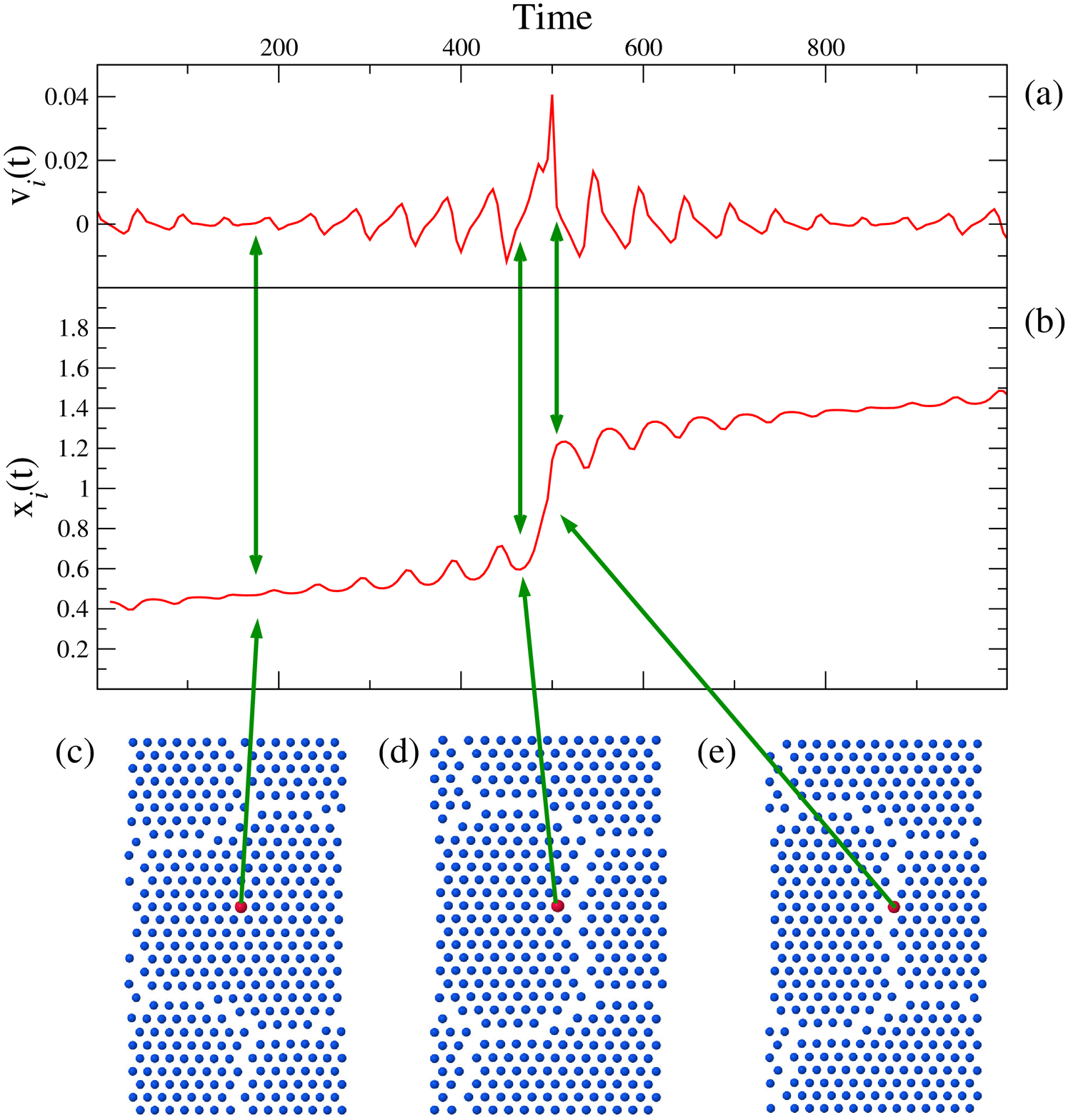}
}
\caption{\label{vel_traj_cell}
  $\rho=14/15$, $T=0$:
  The time evolution of the $x$ component of (a) velocity and (b) position
  of a single (highlighted in panels c-e) colloid, when $F=0.24$ and the CM
  velocity is in the fractional Shapiro step of Fig.~\ref{steps0.93},
  characterized by $v_{\rm cm}/(\nu\,a_{\rm las}) = w = 1/14$.
  Because $\rho < 1$, the forward sliding motion is realized by a backward
  motion of vacancies, here forming antisoliton lines.
  Velocity shows alternating positive and negative spikes, plus a large
  positive peak when the leftward moving domain boundary reaches,
  between $t=465$ and $t=515$, the colloid, which then jumps across the
  antisoliton line, joining the next domain.
  Near the center of the domain -- e.g.\ at $t \simeq 175$, panel (c) --
  the lateral oscillation amplitude reaches its minimum, because the
  particle sits close to a potential well bottom.
}
\end{figure}

Figure~\ref{antisoliton_passage} shows that the antisoliton lines divide
the cell into domains inside which the particles are essentially static
at a minimum of the corrugation potential.
The overall forward sliding motion demanded by the force $F$ is
actuated, for $\rho < 1$, through the backwards motion of the antisolitons,
where all vacancies are lumped.
The antisolitons move to the left in synchrony with the modulation,
advancing faster when the instantaneous corrugation amplitude $V_0 +
\Delta_0 \sin(\omega t)$ of Eq.~\eqref{time_dep_pot} is minimum.
At that moment a fraction of colloids jumps from the left to the right
side of an antisoliton line, while all others remain essentially static.
In the $w\,m/n$ step, this colloid displacement results in the full
antisoliton pattern moving by $m a_{\rm las}$ to the left every $n$
modulation periods.
Figure~\ref{vel_traj_cell} illustrates the dynamics of a single colloid.
Most of the time the particle just oscillates, following the AC
modulation: when the amplitude $V_0 + \Delta_0 \sin(\omega t)$ is
maximum, the particle is driven toward the nearest corrugation-potential
well; when the amplitude is minimum, the particle relaxes due to the
two-body repulsion.
The amplitude of this oscillation is minimal when the colloid is near
the center of a domain, Fig.~\ref{vel_traj_cell}c, and increases as the
colloid reaches the antisoliton, which is the domain boundary.
A positive, uncompensated velocity spike, see
Fig.~\ref{vel_traj_cell}a,b, signals the moment when the particle
crosses the gap and enters the neighboring domain,
Fig.~\ref{vel_traj_cell}d,e.
It should be noted here that some fraction of the colloid rightward
motion also takes place while inside the commensurate domain, which is
not rigid and thus becomes progressively deformed elastically.

\begin{figure}
\centerline{
\includegraphics[width=0.48\textwidth,angle=0,clip=]{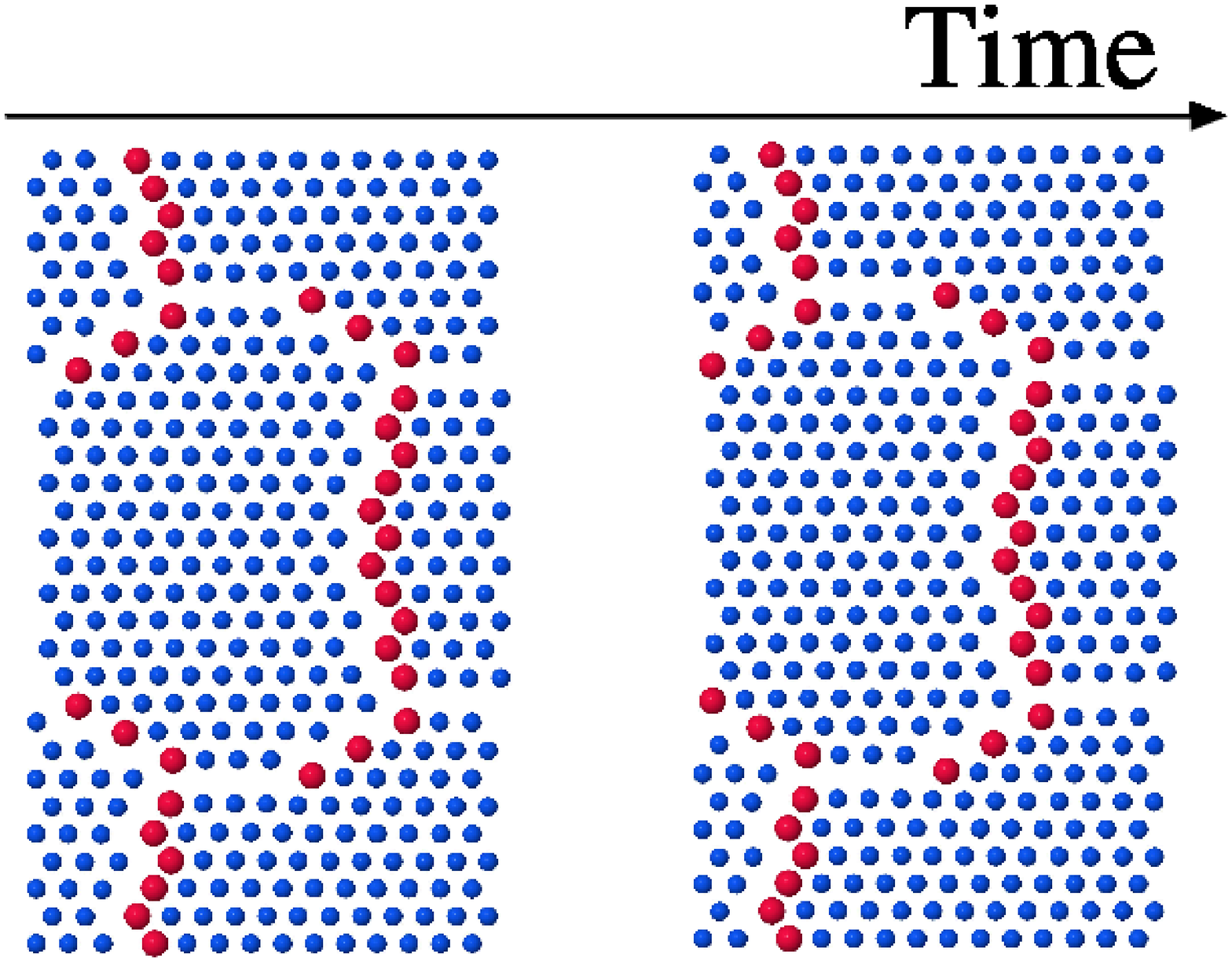}
}
\caption{\label{w_plateau}
  $\rho=14/15$, $N=392$ particles, $V_0 =1$, forward force $F=0.24$,
  step $w=1/14$.
  Successive snapshot patterns of advancing colloids, separated by
  intervals of one oscillation period $\nu^{-1}$.
  The antisolitons drift to the left, causing the CM to move to the
  right.
  The particles whose displacements exceed a fixed threshold $\delta$
  are highlighted (lighter and bigger).
  In this step $22$ particles shift to the right (i.e., along $\hat
  x$) in each oscillation period, crossing the vertical section of the
  antisoliton line, plus another $12$ cross an oblique section, with a
  lateral $\hat y$ component of motion, and only half distance along
  $\hat x$.
  As a result a total of $22+12/2=28$ particles advance by $a_{\rm
    las}$, joining the next commensurate domain.
}
\end{figure}

We are now in a position to clarify the nature of the $w$ subharmonic
steps.
To do this, we observe the colloidal pattern in a 'stroboscopic'
approach, by comparing snapshots taken at successive times separated by
one period, namely when the corrugation amplitude,
Eq.~\eqref{time_dep_pot}, acquires its maximum value $V_0 + \Delta_0$,
i.e.\ for successive half-odd integer values of $\nu t$.
We highlight the particles that during the last period $\nu^{-1}$ have
underwent a displacement exceeding a certain threshold $\delta$.
Due to the elastic domain deformations discussed above this threshold
must be chosen appropriately: for the conditions considered here a value
$0.12\,a_{\rm coll}\le \delta\le 0.61\,a_{\rm coll}$ provides consistent
results for all plateaus.
The highlighted particles represent colloids, whose number we designate
as $m_i$, which cross the antisoliton boundary between two neighboring
domains in that period $i$ of oscillation.
In the fundamental plateau where the quantized speed equals $w$, the
pattern of advancing particles is the same at all oscillation periods,
as seen, e.g., in Fig.~\ref{w_plateau} and equals $Nw$ where $N$ is the
total number of particles.
In the present case of Fig.~\ref{w_plateau} $N= 392$, $w = 1/14$, the
Shapiro plateau has dimensionless velocity $w$, and a total of $Nw=28$
particles cross the boundary to the next domain at each oscillation
period.
More specifically, $22$ colloids at the edge of the island aligned along
$\hat y$ move straight in the $\hat x$ direction into the next
corrugation minimum at a distance $a_{\rm las}$.
In addition, $12$ more particles along the oblique antisoliton line
advance with a sidewise $\hat y$ component, thus proceeding into a
minimum which is $a_{\rm las}/2$ to the right.
The total movement amounts therefore to effectively $22+12/2=28 =Nw$
particles advancing by $a_{\rm las}$ in each period.
The very same picture applies to sliding a monolayer with any different
lattice-spacing ratio $\rho=(1+w)^{-1}$.
The fundamental subharmonic step occurs again at velocity $w$.
%

\begin{figure}
\centerline{
\includegraphics[width=0.48\textwidth,angle=0,clip=]{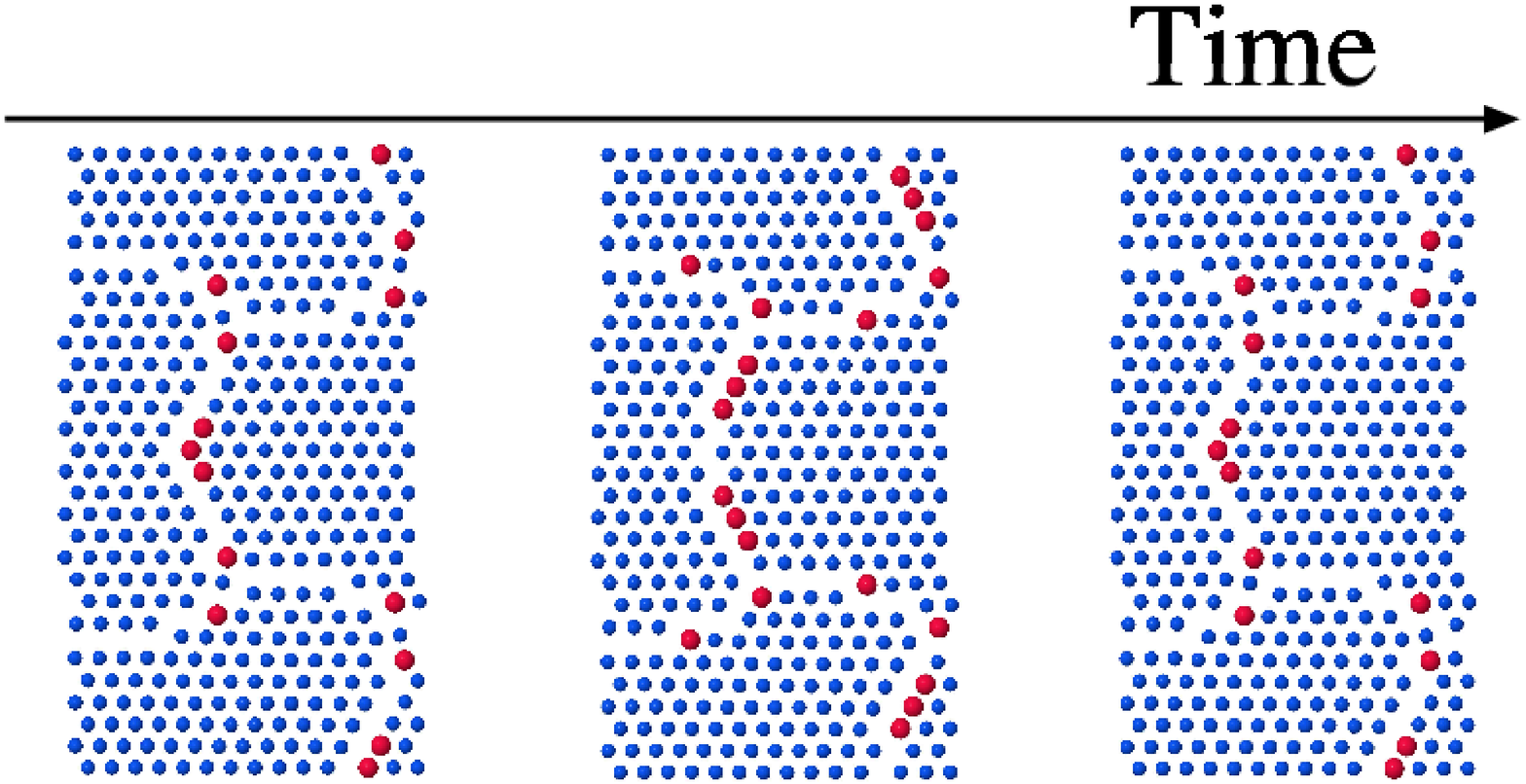}
}
\caption{\label{w_2_plateau}
  The subharmonic plateau $w/2$ for $\rho=14/15$, $w=1/14$, $F=0.185$.
  Successive patterns of advancing colloids, separated by intervals of
  one oscillation period $\nu^{-1}$, illustrating the propagation of an
  antisoliton in this Shapiro plateau.
  In a first displacement $10+4/2=12$ (highlighted) particles jump (the
  $4$ particles crossing oblique antisolitons advance by a half lattice
  spacing along $\hat x$).
  Then, at the next period, another $12+8/2=16$ (highlighted) particles
  complete the advancement of the string of particles bordering the
  antisoliton line.
  At successive periods, the alternating advancement of $12$ and $16$
  particles repeats itself involving successive lines of colloids more
  and more to the left.
}
\end{figure}

The pattern of advancing colloids is somewhat more complex in the other
subharmonic steps.
We explore these differences in Figs.~\ref{w_plateau} and
\ref{w_2_plateau}.
To exemplify, let us consider again $w=1/14$, but now focus on the $w/2$
plateau, Fig.~\ref{w_2_plateau}.
Here, a total particle number $Nw$ again jumps between a domain and the
next: but that takes place in {\it two} successive oscillation periods
instead of one as in the harmonic $w$ step.
In the first AC oscillation period $10+4/2=12$ colloids cross the
antisoliton line, whereas the remaining $12+8/2=16$ jump in the second
period.
Similarly, in the step leading to the $w/3$ plateau, it takes a sequence
of three successive oscillation periods to complete the advancement
which transfers $Nw$ particles across the antisoliton.
The partial transfers are in this case $10+4/2=12$, $8+4/2=10$, and $4+4/2=6$
particles in period 1, 2 and 3 respectively.
%

In summary, each Shapiro step is characterized by a periodic advancement
pattern during which an integer multiple of $Nw$ particles jump from a
commensurate domain to the next.
However the $m/n$-subharmonic step of quantized velocity
\begin{equation}\label{wovern_eq} 
 \frac {v_{\rm cm}}{\nu\,a_{\rm las}} = w\, \frac mn
\end{equation} 
is accomplished as a composite of $n$ successive modulation periods,
each lasting $\nu^{-1}$, of the AC modulation.
After one such complete migration, the entire pattern of positions is
displaced bodily by $m$ lattice spacings $a_{\rm las}$.
The subharmonic step leading to $w\,m/n$ is thus composite, involving $n$
AC modulation periods.
The total number $mNw$ of particles that cross the antisoliton is
partitioned between the $n$ periods,
\begin{equation}\label{partition} 
mNw = \sum_{i=1}^n  m_i
 \,.
\end{equation} 
The detailed numbers $m_i$ of crossing particles at each individual AC
period $i$ will generally depend on $\rho$, $V_0$, $\Delta_0$, and on
the angle between the sliding direction and the antisoliton (soliton)
orientation.
Occasionally, the partition \eqref{partition} may even change along a
subharmonic step as $F$ increases.
We have not found a general analytic rule, if any, determining this
partition.

\subsubsection{Effect of the corrugation and modulation amplitudes.}
\label{Amplitude}

We now investigate the role played by the corrugation amplitude $V_0$
and by the modulation amplitude $\Delta_0$ in the formation of
Shapiro-step structures as exemplified in the $\rho=14/15$ mismatched
configuration.
First of all we remind ourselves that below the pinning threshold
corrugation amplitude $V_{0\, \rm crit}$ (which, as seen on
Fig.~\ref{aubry_explanation} is around $0.4$ for the present case
$\rho=14/15$) the layer is superlubric, i.e., there is no static
friction (other than that, negligible, due to the weak $14/15$
commensurability), and a time dependent modulation with amplitude
$\Delta_0$ such that $V_0+\Delta_0<V_{0\, \rm crit}$ would produce no
Shapiro steps.
Thus we concentrate on corrugation amplitudes above the threshold, where
the particle lattice is strongly pinned.

\begin{figure}	
\centerline{
\includegraphics[width=0.48\textwidth,angle=0,clip=]{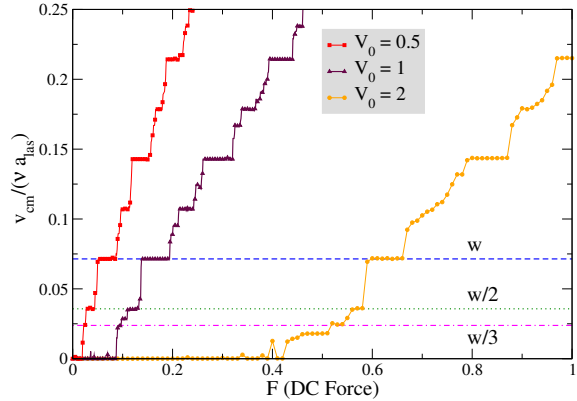}
}
\caption{\label{corrugation_compar}
  $\rho=14/15$: rescaled CM velocity for several corrugation amplitudes
  $V_0$, all with a modulation amplitude $\Delta_0 = 70 \%\,V_0$ and
  frequency $\nu = 0.02$.
  The initial depinning force $F$ increases rapidly with the corrugation
  amplitude $V_0$.
  Shapiro steps widen and shift to larger forces when the corrugation is
  increased.
}
\end{figure}

Figure~\ref{corrugation_compar} is useful to understand the corrugation
dependence of Shapiro steps.
We probe three corrugation values: $V_0=0.5$, $V_0=1$ and $V_0=2$ all
with an amplitude modulation $\Delta_0=70\%V_0$.
The whole plateau structure moves to larger forces when the corrugation
$V_0$ is increased.
As expected from common sense, the depinning force increases
systematically with growing corrugation $V_0$, nearly $\sim 20$ times
larger when $V_0$ changes from $0.5$ to $2$.
With reference to Fig.~\ref{aubry_explanation}, both $V_0=0.5$ and
$V_0=1$ have $V_0-\Delta_0 < V_{0\,\rm crit}$, thus depinning occurs at
an arbitrarily small force, also depending on frequency as discussed
below in Sect.~\ref{depinning:sec}.
In contrast, $V_0=2$ has $V_0-\Delta_0=0.6>V_{0\,\rm crit}$, with a
broad range of forces for which, independently of frequency, the system
remains pinned, thus explaining the huge increase of static friction.
The width of the Shapiro steps, especially the smaller ones, also
increases with $V_0$: for example, the $w/3$ step, barely detectable for
$V_0 = 0.5$ and $1$, becomes much broader for $V_0= 2$.
In general, a larger potential $V_0$ amplitude, corresponding to a
comparably softer colloid lattice, spatially narrower solitons, and
larger static friction, enhances the Shapiro-step structure, which
emerges more clearly.

\begin{figure}
\centerline{
\includegraphics[width=0.48\textwidth,angle=0,clip=]{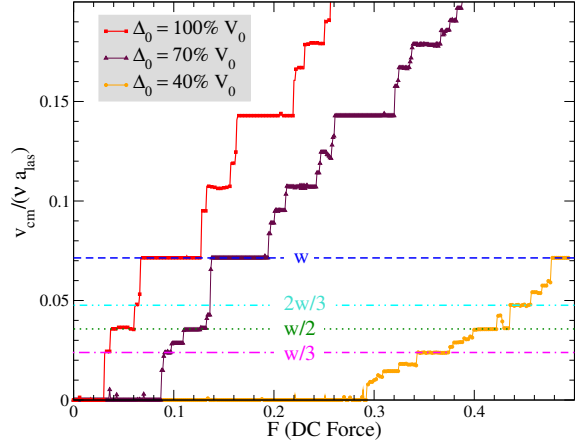}
}
\caption{\label{modulation_compar}
  Velocity of the colloid CM (rescaled as in
  Fig.~\ref{shapiro_commensurate}), for $\rho=14/15$, $T=0$, oscillation
  frequency $\nu=0.02$, corrugation amplitude $V_0=1$, for different AC
  oscillation amplitudes: $\Delta_0=40\%\,V_0$ (dots),
  $\Delta_0=70\%\,V_0$ (triangles), $\Delta_0=100\%\,V_0$ (squares).
  Note that the smaller modulation $\Delta_0 = 40 \%\, V_0$ exhibits a
  richer pattern of steps.
  The width of the main subharmonic steps tends to increase with increasing
  modulation amplitude $\Delta_0$, at the expense of minor steps.
}
\end{figure}

Next, we explore the AC modulation amplitude dependence,
Fig.~\ref{modulation_compar}, adopting $V_0=1$ and varying $\Delta_0$
from $40\% \,V_0$, to $\Delta_0=100\% \,V_0$ (the purple triangles are
indeed the same data as the purple triangles of
Fig.~\ref{corrugation_compar}).
As in the lattice-matched case Fig.~\ref{comm_V0.5}, when $\Delta_0$
increases, the Shapiro steps become wider and extend further down to
smaller force $F$.
Moreover, a larger AC modulation $\Delta_0$ depresses large-$n$
longer-period subharmonic $w/n$ in favor of shorter ones with smaller
$n$ (e.g $w/2$ or $w$).
Conversely, as Fig.~\ref{modulation_compar} shows, all individual
plateaus shrink and become more numerous for smaller modulation
amplitudes.
Moreover, with $\Delta_0=40\%\,V_0$ we detect, in addition to $w$,
$w/2$, $ w/3$, and $w/5$, extra fractional plateaus such as $4w/5$,
$2w/3$.
In short, a small AC modulation amplitude $\Delta_0/V_0$ produces many
weak subharmonic steps, while a large modulation promotes fewer stronger
and wider ones.

\subsection{$\rho=3/4$}

\subsubsection{Step structure.}

\begin{figure}
\centerline{
\includegraphics[width=0.48\textwidth,angle=0,clip=]{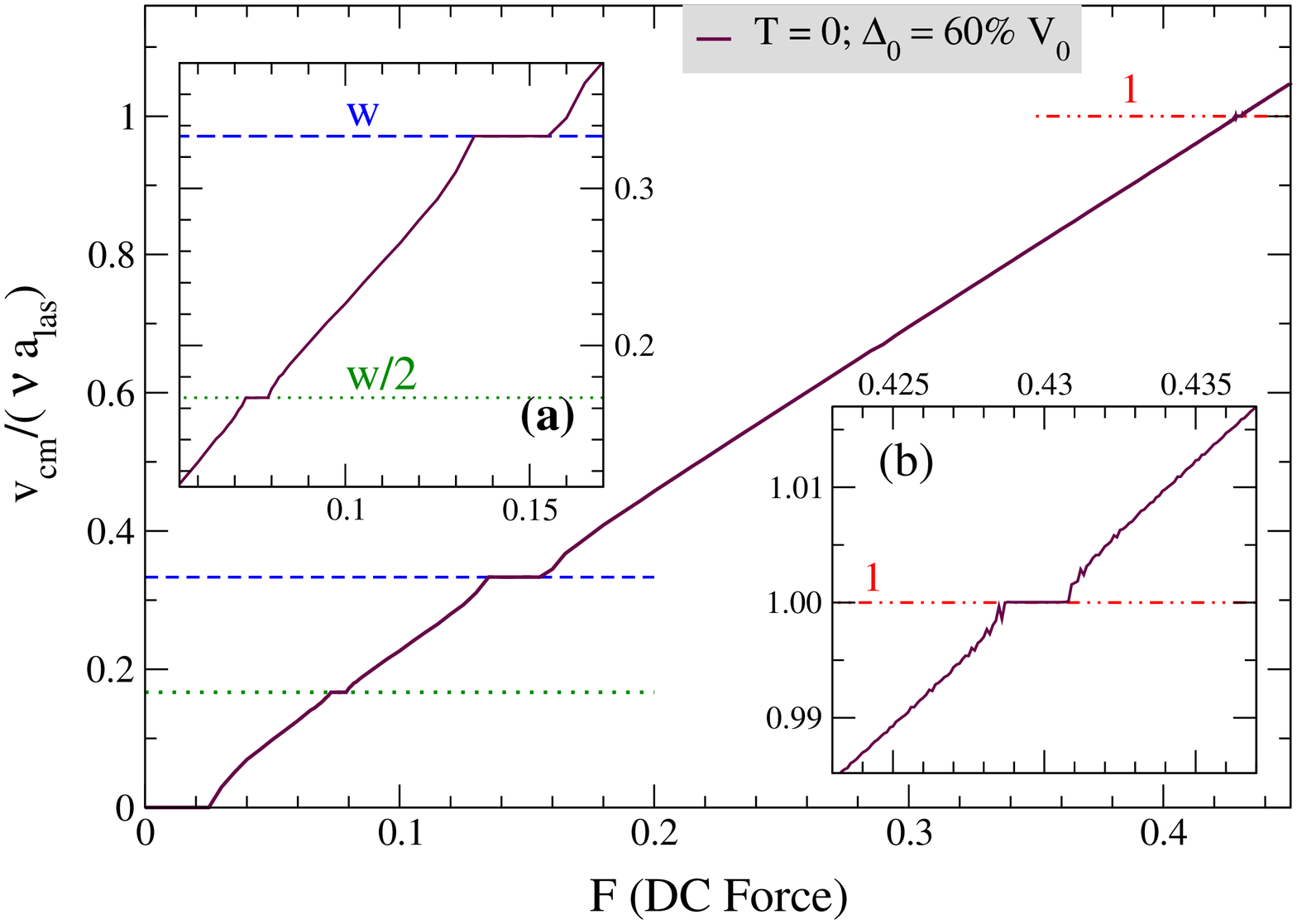}
}
\caption{\label{t_0_f_0.75_V0_1_D0_60}
  $\rho=3/4$, $w = 1/3$, $N= 450$ particles, corrugation $V_0= 1$,
  modulation $\Delta_0 = 60\% \,V_0$, oscillation frequency $\nu =
  0.02$. Two subharmonic Shapiro steps are visible.
  Inset (a): Magnified view of the subharmonic plateaus.
  Inset (b): Magnification of the ``trivial'' plateau, $v_{\rm cm}
  /(\nu\,a_{\rm las}) = 1$ representing all colloids advancing as one
  every modulation period.
}
\end{figure}

There is a clear contrast between the "nearly incommensurate" case, such
as $\rho= 14/15$, or $w=1/14$, and a "strongly commensurate" one, such
as $\rho=3/4$, corresponding to $w=1/3$, Eq.~\eqref{w_eq}.
In this strongly-commensurate case the Shapiro-step structure --
Fig.~\ref{t_0_f_0.75_V0_1_D0_60} -- shows fewer steps than in the nearly
incommensurate case of Fig.~\ref{steps0.93}.
Basically only two subharmonic plateaus ($w$ and $w/2$) plus the
``trivial'' step are detected.
Moreover, the width of these plateaus is smaller than it was for $\rho=
14/15$.
However, as was the case there, even for strong commensurability
$\rho=3/4$ the plateau widths increase for increasing corrugation
amplitude, and that even more rapidly than for $\rho=14/15$.

\subsubsection{Subharmonic advancement of antisolitons.}

\begin{figure}
\centerline{
\includegraphics[width=0.48\textwidth,angle=0,clip=]{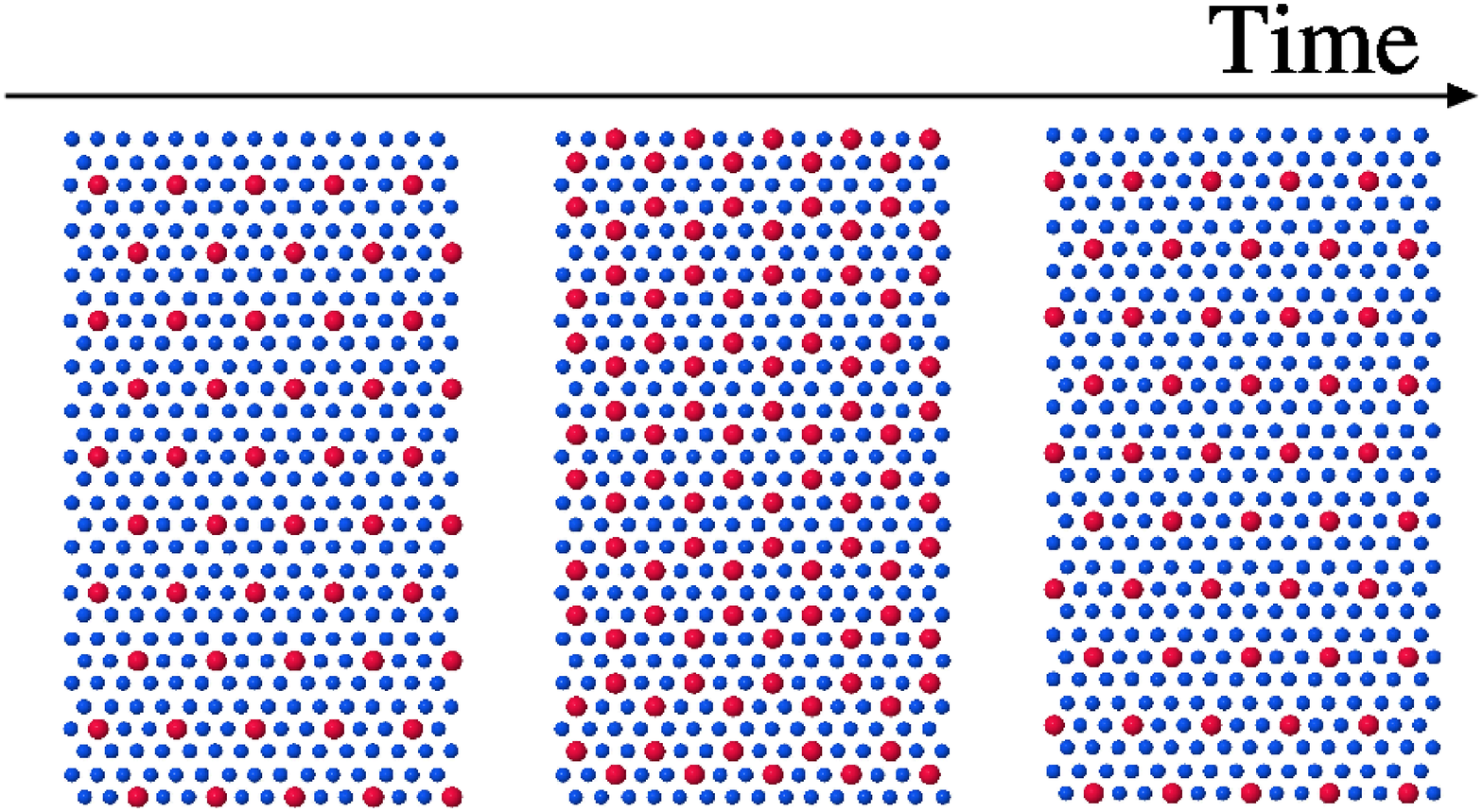}
}
\caption{\label{3_4_w_2}
  A stroboscopic picture of the $\rho=3/4$ advancement pattern, similar to
  Fig.~\ref{w_2_plateau}, for the $w/2$ step.
  The antisoliton pattern is so weak that individual hexagonal domains
  are nearly invisible.
  The advancing colloids (bolder particles) are highlighted: $150$
  particles advance in two sets at successive modulation periods:
  alternately $50$ (left), then $100$ (center), then again $50$ (right),
  and so on.
}
\end{figure}

We examine the advancement mechanism for the subharmonic plateaus in
Figure~\ref{3_4_w_2}.
The mechanism is again similar to that of Sec.~\ref{14_15},
Fig.~\ref{w_2_plateau}.
Since $\rho=3/4$ is further away from unity than $14/15$, the
antisoliton pattern is composed by much closer, partly overlapping and
less distinct lines.
The subdivision in domains is nearly invisible, with each particle
always approximately as close to the edge of a domain as to its center.
As a result, the oscillations around the average colloid trajectory are
far less evident than in Fig.~\ref{vel_traj_cell}.
Nevertheless, the advancement mechanics is the same as in the earlier
case: e.g.\ the $w/n$ subharmonic step involves $Nw =150$ particles,
advancing during $n=2$ periods: we identify a clear pattern of advancing
particles, displayed in Fig.~\ref{3_4_w_2}.
Here $150$ colloids at the domain boundary advance in a first group of
$50$ particles in one modulation period, then a second group of $100$ in
the next period.
In the $w$ step, all $150$ particles at the boundary cross the
antisoliton line at every period: they could be identified by
superimposing the highlighted colloids in the first and second panels of
Fig.~\ref{3_4_w_2}.

\subsubsection{Thermal effects.}
\label{Thermal effects}

Finally we explore thermal effects on the subharmonic step structure.
All tests in the $\rho=14/15$ system show that at finite temperature the
fragile Shapiro structure is strongly affected: the steps are washed
away by the colloidal diffusive motion.
No clear step can be detected for $T > 0.001 \ll T_{\rm room} = 0.04$,
whether the standard $V_0=1$ and $\Delta_0= 60\%\, V_0$ or other
parameters are adopted.

\begin{figure}
\centerline{
\includegraphics[width=0.48\textwidth,angle=0,clip=]{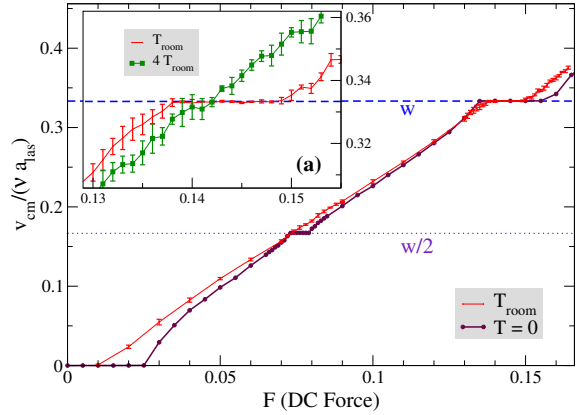}
}
\caption{\label{t_0.04_f_0.75_V0_1_D0_60}
  Thermal effects in the $\rho=3/4$ mismatched configuration, for
  $V_0=1$, $\Delta_0=60\%\,V_0$.
  Like in Fig.~\ref{T_comm_0.04}, the mean value and error bar of
  $v_{\rm cm}$ are computed by averaging over several simulations.
  The $w$ step remains quite well identified at $T_{\rm room}$, 
  while the $w/2$ step disappears altogether.
  (a): a zoom-in of the subharmonic $w$ step at $T= T_{\rm room}$
  compared with the much larger temperature $T = 4\,T_{\rm room}$ where
  the plateau structure is washed out by thermal noise.
}
\end{figure}

By contrast, with the $\rho=3/4$ mismatch ratio, where commensurability
is stronger, the effects of thermal fluctuations are less dramatic, and
at least the strongest steps survive.
As shown in Fig.~\ref{t_0.04_f_0.75_V0_1_D0_60}, the $w$ plateau persists up
to and above $T_{\rm room}$, the thermal effects leading to a slight
reduction in width and to a slight edge rounding compared to $T=0$.
We explored even higher temperature, and found that even at $T =
4\,T_{\rm room}$ the $w$ step leaves some remnant of its existence in
the force-velocity diagram, Fig.~\ref{t_0.04_f_0.75_V0_1_D0_60}a.
However, even room-temperature fluctuations are enough to cancel out
completely the $w/2$ plateau.

\begin{figure}
\centerline{
\includegraphics[width=0.48\textwidth,angle=0,clip=]{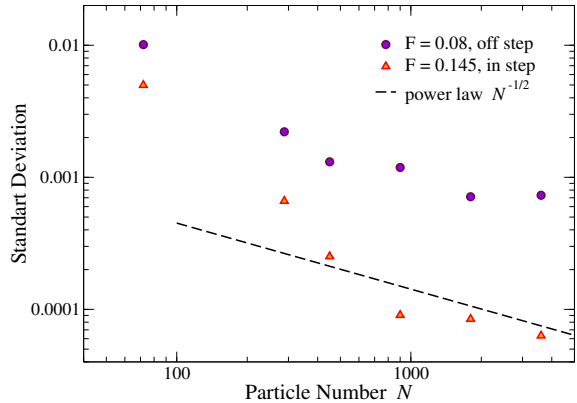}
}
\caption{\label{error_dep}
  Size scaling of the fluctuations of the CM velocity in
  room-temperature simulations of the $\rho=14/15$ mismatched
  configuration for $V_0=1$, $\Delta_0=60\%\,V_0$, executed for
  $F=0.08$ (climb between steps) and $F=0.145$ (step $w$) of
  Fig.~\ref{t_0.04_f_0.75_V0_1_D0_60}.
  The various sizes are simulated with rectangular cells similar to the
  one depicted in Fig.~\ref{simulation_cell}.
  The fluctuations are obtained as the standard deviation evaluated over
  4 independent simulations for each size.
  Dashed line: a $N^{-1/2}$ guide to the eye.
}
\end{figure}

Thermal effects on Shapiro steps also depends to some extent on size.
As long as the colloid density remains uniform, thermal fluctuations
average out for increasing sample size and averaging time: for a given
temperature, a larger sample size and longer simulation show flatter,
sharper plateaus.
Figure~\ref{error_dep} shows how the thermal fluctuations on the colloid
CM speed depend on the number $N$ of colloids in the simulation
supercell.
For large size, as expected by statistics, thermal fluctuations scale as
$N^{-1/2}$, while when the cell lateral size is comparable to or smaller
than the correlation length associated to the colloid-colloid
interaction, dynamical correlations enhance the fluctuations.
Note that inside a plateau (triangles in Fig.~\ref{error_dep}), the
fluctuations are much smaller and the correlation length (marked by the
knee in the size scaling) significantly longer, than in a climb between
steps.

\section{The depinning force}\label{depinning:sec}

\begin{figure}
\centerline{
\includegraphics[width=0.48\textwidth,angle=0,clip=]{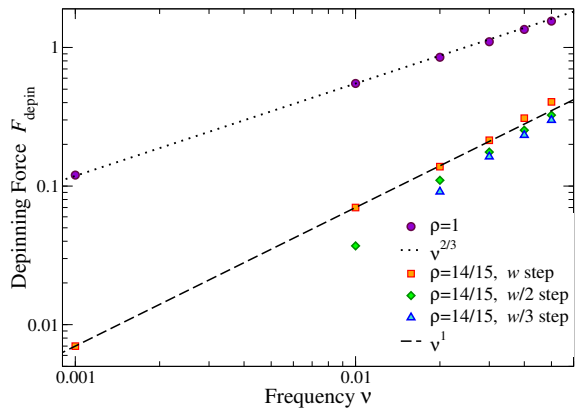}
}
\caption{\label{depinning:fig}
  The static depinning force as a function of the modulation frequency
  in conditions where the sliding barrier vanishes during the modulation
  period.
  Circles: the lattice-matched geometry, with $V_0=1$, $\Delta_0=100\%\,V_0$.
  Other symbols: $\rho=14/15$, with $V_0=1$, $\Delta_0=70\%\,V_0$.
}
\end{figure}

This section deals with the DC-Force -- corrugation-amplitude diagram,
Fig.~\ref{aubry_explanation}.
We can foresee that, depending on $V_0$ and $\Delta_0$, the static-friction
force $F_{\rm depin}$, namely the minimum force required to maintain a net
nonzero sliding speed, can fall in one of two alternative regimes.
The simplest condition occurs when $V_0-\Delta_0$ exceeds the Aubry
$V_{0\,\rm crit}$: then the colloid remains statically pinned throughout
the modulation period, with the result that sliding requires a finite
driving force greater or equal than the depinning force appropriate to a
corrugation $V_0-\Delta_0$ in the absence of modulation.

Conversely, when $V_0-\Delta_0 < V_{0\,\rm crit}$, for a fraction
$f_{\rm free}$ of the oscillation period $\nu^{-1}$, the pinning barrier
vanishes, and the colloid slides freely.
A slow enough AC modulation gives the particles enough time to take
advantage of the temporary lack of pinning and generate the first
Shapiro step for arbitrarily small driving force $F$.
In such condition, the depinning force is a function of the modulation
frequency.
The depinning force $F_{\rm depin}$ can be estimated as follows:
During the unpinned time $f_{\rm free}\nu^{-1}$ the colloids advance at
a speed $\simeq \mu F$, where $\mu\simeq \eta^{-1}$ is the colloid
mobility; by equating the displacement $f_{\rm free}\nu^{-1} F \mu$ to a
minimum advancement compatible with the overall displacement in the
lowest Shapiro step, say
$\frac wn a_{\rm las}$, 
we obtain
\begin{equation}\label{depinning_linear}
  F_{\rm depin} \simeq
  \frac{w a_{\rm las}}{n \mu f_{\rm free}}\,\nu
 \,.
\end{equation}
Thus, under modulation conditions such that $V_0-\Delta_0 < V_{0\,\rm
  crit}$, the static friction force is predicted to vanish linearly with
the modulation frequency.
Simulations confirm this power law scaling, indicated by the dashed line
in Fig.~\ref{depinning:fig}.
For decreasing frequency, simulations also show that the $n>1$
subharmonic steps tend to weaken and eventually disappear, see e.g.\ the
diamonds ($w/2$) and triangles ($w/3$) in Fig.~\ref{depinning:fig}.
As a result, eventually at very low frequency depinning occurs directly
into the $w$ step.

The fully-matched case $\rho=1$ discussed in Sec.~\ref{commensurate:sec}
has no Aubry transition, so $V_{0\,\rm crit}=0$.
Even so, at $100 \%$ modulation rate ($\Delta_0=V_0$), in the brief
instants when $\sin(\omega t)\approx -1$, again free sliding is allowed.
At finite driving force $F$ a finite time interval $f_{\rm
  free}\nu^{-1}$ is compatible with barrier overcoming and consequent
free sliding.
This interval is obtained by evaluating the region of $F\geq F_{s1}$
using the expression \eqref{F_s1} (under the condition $\Delta_0=V_0$)
for the barrier crossing force, obtaining
\begin{equation}\label{f_free_comm}
  f_{\rm free} =
  \frac 12 + \frac 1{\pi}
  \arcsin\!\left(\frac{9 F a_{\rm las}}{8\pi V_0} -1\right)
 .
\end{equation}
By expanding the $\arcsin$ function with its argument near $-1$ (small
driving force) we obtain
$f_{\rm free}\simeq [9 F  a_{\rm las}/(4 \pi^3 V_0)]^{1/2}$.
As above, we then equate the displacement $f_{\rm free}\nu^{-1} F \mu$
to the minimum advancement required for the colloids to advance by one
corrugation lattice spacing, say $\frac 12 a_{\rm las}$, obtaining
\begin{equation}\label{depinning_23}
  F_{\rm depin} \simeq
  \pi \left(\frac {V_0 a_{\rm las}}{9 \mu^2}\right)^{\!1/3} \nu^{2/3}
 \,.
\end{equation}
Again, simulations confirm this result, see Fig.~\ref{depinning:fig}, with
depinning always occurring into the trivial harmonic step.
In principle a similar ``critical'' $\nu^{2/3}$ power-law regime is to
be expected even in the mismatched case $\rho<1$, under the condition
$V_0-\Delta_0 = V_{0\,\rm crit}$ of an instantaneous touching of the
free-sliding superlubric Aubry phase during each modulation cycle.

\section{Discussion and Conclusion}\label{Discussion and Conclusion}

The present work addresses, by means of MD simulations, the question
whether Shapiro step structures can occur in the forced sliding of a 2D
colloidal monolayer immersed in an optical lattice spatially periodic
corrugation potential whose amplitude has a static DC component plus an
AC modulation that oscillates sinusoidally in time.
We confirm that Shapiro steps should appear, in the form of intervals of
static force within which the CM colloid velocity is quantized, at least
at zero temperature, to a value entirely determined by the modulation
frequency and by geometric parameters.
The only condition required for this to occur at $T=0$ is that in the
absence of the AC modulation the colloids should be pinned by static
friction, and not free sliding, as can happen when the 2D colloid
lattice and the optical lattice are mismatched and the optical lattice
is sufficiently strong.
In commensurate, perfectly lattice-matched conditions $\rho=1$ only
"trivial" integer steps are expected, corresponding to the synchronized
periodic advancement of all particles at each AC modulation period.
In lattice-mismatched conditions $\rho \neq 1$ we find additional
subharmonic Shapiro plateaus beside the integer ones.
The crucial quantity governing the subharmonic plateaus is $w =
|1-\rho^{-1}|$, which determines the quantized velocity step value in
the totally general form
\begin{equation}\label{cms-speed} 
v_{cm} = \nu \, a_{\rm las} w\,\frac mn
 \,,
\end{equation} 
where $\nu$ is the AC frequency, $a_{\rm las}$ is the optical lattice
spacing, and $m$ and $n$ are small integers.
Interestingly, we find that every quantized Shapiro step has a specific
real-space signature corresponding to which particles advance
synchronously with the AC modulation -- a signature which is purely
geometric.
In a step the total number of advancing particles per modulation period
equals exactly a fraction $N\,w\,m/n$ of the total particle number $N$.
However, a step corresponding to integer $n$ lasts $n$ periods.
During each period $\nu^{-1}$ only a sub-fraction of these particles
advance by one lattice step, and the total $Nwm$ is achieved only at the
end of a cycle of $n$ periods.

We analyzed how the corrugation amplitude $V_0$ and modulation $\Delta_0$
influence the ensuing step structure.
An increase in $V_0$ leads to a comparably softer colloidal layer, thus
increased localization and static friction, resulting in a shift to
higher forces $F$ of the Shapiro plateaus and generally an increase in their
width.
On the other hand an increase in $\Delta_0$ to the point of making
$V_0-\Delta_0<V_{0\, \rm crit}$ leads the oscillation to
explore more and more of the hard-colloidal-layer region for part of the
oscillation period.
In this region, free sliding is possible, as well know in the 1D FK model
\cite{Peyrard83}.
This results in more extended steps, in particular down to smaller $F$.
In such cases, the smallest (depinning) force is a function of the
modulation frequency, and we verified that it changes either as $\nu$ as
$\nu^{2/3}$ depending on whether a free sliding state is crossed for a
finite fraction of the modulation period or just touched instantly.
Additionally, a very large $\Delta_0 \simeq V_0$ tends to simplify the
rich ladder of subharmonic step structures observed at intermediate
$\Delta_0$.

We also checked how important the specific shape of the
corrugation-amplitude time modulation really is for the possibility to
observe Shapiro steps.
We ran simulations both for the lattice-matched case and for the
$\rho=3/4$ mismatch ratio, replacing $\sin(\omega t)$ in
Eq.~\eqref{time_dep_pot} with a square wave ${\rm sign}(\sin(\omega
t))$, where ${\rm sign}(x)$ equals $x/|x|$ for nonzero $x$.
Compared to the sinusoidal case with identical modulation amplitude
$\Delta_0$ and other conditions, the square-wave simulations show quite
similar velocity-force profiles, with marginally wider and more robust
Shapiro steps.
This small difference can be understood as a consequence of the larger RMS
modulation of the square wave than the sinusoidal wave for the same
peak-to-peak amplitude.

Thermal fluctuations and density inhomogeneities present in real colloid
sliding experiments are likely to affect the possibility to observe
Shapiro structures.
While the perfectly-matched system is barely affected by the thermal
Brownian motion, the subharmonic Shapiro-step structures of mismatched
systems are far more fragile.
Near matching between the colloid and the optical lattice, the loose
soliton network is strongly affected by thermal fluctuations, which wash
easily away most of the delicate Shapiro plateaus.
By contrast, for a more robust (but commensurate) lattice mismatch such as
$\rho=3/4$, we retrieve subharmonic steps up to $T_{\rm room}$ and even
above.
We test also the size dependence in the latter case: as long as the
density is uniform, a greater number of particles averages out thermal
fluctuation, ending up in a flatter and less rounded step.

The model explored in the present work neglects hydrodynamic
interactions \cite{Diamant05}: while such velocity-dependent many-body
forces may affect the quantitative detail of the Shapiro steps, we do
not expect they would change the qualitative picture much.
In particular, nonuniform solitonic motion, as occurs in mismatched
configurations, could be enhanced by hydrodynamic interactions.
We are currently investigating this issue in quantitative detail.
Also, the detail of the two-body interaction,
Eq.~\eqref{coulomb-yukawa}, is not likely to play any qualitative role
in the Shapiro-steps physics.
In particular, we expect a similar behavior if the colloid-colloid
interaction, rather than screened electrostatic, was the power law
describing the repulsion between magnetic colloidal particles in a
layer, controlled by a perpendicular magnetic field
\cite{Bizdoaca01,Xu02,Ge08}.

In conclusion, this theoretical study predicts that Shapiro-like plateau
structures of colloid CM velocities should be easily detected
experimentally in a 2D rich geometry-- at least in lattice-matched
condition, with potential amplitude and oscillation modulation values
varying over broad ranges.
In mismatched conditions, on the other hand, the experimental detection
of subharmonic plateaus seems limited by the ability to provide large
regions where the 2D colloid density is constant in space with
sufficient accuracy.
While experimentally this may not be possible for all subharmonic
quantizations, still for a lattice-spacing ratio sufficiently distant
from unity, a step structure should be detectable under experimental
conditions.
%
Subharmonic steps were indeed detected (although not identified as such)
in the simpler 1D experimental setup of Ref.~\cite{Juniper15}.

\section*{Acknowledgments}

We acknowledge useful discussion with Clemens Bechinger.
This work was partly supported under the ERC Advanced Grant
No.\ 320796-MODPHYSFRICT, by the Swiss National Science Foundation
through a SINERGIA contract CRSII2\_136287, by PRIN/COFIN Contract
2010LLKJBX 004, and by COST Action MP1303.

\section*{References}


\begin{thebibliography}{10}
\bibitem{AubryRational:note}
{ The Aubry transition occurs rigorously only for strictly incommensurate
  (irrational length ratio $f$) conditions. In rationally matched situations
  such as those considered here, the ``superlubric'' state is characterized by
  an extremely small nonzero static friction, which vanishes exponentially with
  the distance from the Aubry point $V_{0\, \rm crit}$}.
\end{thebibliography}

\end{document}